%%%%%%begin TeX source
%\magnification=\magstep1
%\baselineskip=22pt
\def\etal{{\it et al.}}
%\centerline{\bf The Case for Inertia as a Vacuum Effect:
% A Reply to Woodward and Mahood}
\centerline{\bf THE CASE FOR INERTIA AS A VACUUM EFFECT:}
\centerline{\bf A REPLY TO WOODWARD AND MAHOOD}
\bigskip
\centerline{York Dobyns}
\centerline{C-131 Engineering Quad, Princeton University}
\centerline{Princeton, NJ 08544-5263}
\bigskip
\centerline{Alfonso Rueda}
\centerline{Department of Electrical Engineering, ECS-561}
\centerline{California State University, Long Beach, CA 90840}
\bigskip
\centerline{Bernard Haisch$^a$}
\footnote{}
{\noindent $^a$current address: California Institute for Physics and
Astrophysics, 366 Cambridge Ave., Palo Alto, CA 94306, www.calphysics.org}
\centerline{Solar \& Astrophysics Laboratory}
\centerline{Dept. L9-41, Bldg. 252, Lockheed Martin}
\centerline{3251 Hanover St., Palo Alto, CA 94304}
\bigskip

\centerline{\it Foundations of Physics, in press, Jan. 2000}

\bigskip
\noindent{\bf ABSTRACT: }
The possibility of an extrinsic origin for inertial reaction forces has
recently seen increased attention in the physical literature. Among theories
of extrinsic inertia, the two considered by the current work are (1) the
hypothesis that inertia is a result of gravitational interactions, and
(2) the hypothesis that inertial reaction forces arise from the interaction
of material particles with local fluctuations of the quantum vacuum.
A recent article supporting the former
and criticizing the latter is shown to contain substantial errors.

\bigskip\noindent{\bf 1. INTRODUCTION}
\bigskip\noindent
Since the publication of Newton's {\it Principia} the default assumption
of most physicists has been that inertia is intrinsic to mass. Theories
of an extrinsic origin for inertia, however, have seen perennial if 
minor interest. Since the task of physics is to explore causative
relationships among natural phenomena, it is appropriate for physicists
to devote some work to asking how and why the property of mass arises to
produce the phenomenon of  inertia, rather than always and only treating
it as a definitional property.  Recent work, on the other hand, provides a
more urgent reason to look into theories of extrinsic inertia: some of them
suggest a resolution to one of the more intractable difficulties of current
physical theory.

There appears to be a fundamental conflict between quantum theory and 
gravitational theory. Adler, Casey, and Jacob$^{(1)}$ have dubbed this the 
``vacuum catastrophe" to parallel the ``ultraviolet catastrophe" associated 
with blackbody radiation 100 years ago. Quantum field theory predicts a very
large vacuum zero-point energy density, which according to general
relativity theory (GRT) should have a huge gravitational effect. The
discrepancy between theory and observation may be 120 orders of magnitude.
As Adler
\etal\ point out: ``One must conclude that there is a deep-seated
inconsistency between the basic tenets of quantum field theory and gravity."

The problem is so fundamental that elementary quantum mechanics suffices
to demonstrate its origin. The intensity of any physical field, such as
the electromagnetic field, is associated with an energy density; therefore
the average field intensity over some small volume is associated with 
a total energy. The Heisenberg uncertainty relation (in the $\Delta E
\Delta t$ form) requires that this total energy be uncertain, in inverse
proportion to the length of time over which it obtains. This uncertainty
requires fluctuations in the field intensity, from one such small volume
to another, and from one increment of time to the next; fluctuations which
must entail fluctuations in the fields themselves, which must be seen to
be more intense as the spatial and temporal resolution increases. 

In the more formal and rigorous approach of quantum field theory,
the quantization of the electromagnetic field is
done ``by the association of a quantum-mechanical harmonic oscillator with
each mode $\ldots$ of the radiation field."$^{(2)}$
Application of the Heisenberg uncertainty relation to a harmonic
oscillator immediately requires that its ground state
have a non-zero energy of $\hbar\omega/2$, because a particle cannot
simultaneously be exactly at the bottom of its potential well and have
exactly zero momentum. The harmonic oscillators of the EM field are
formally identical to those derived for a particle in a suitable potential
well; thus there is
the same $\hbar\omega/2$ zero-point energy expression for each mode of the
field as is the case for a mechanical oscillator.  Summing up the energy
over the modes for all frequencies, directions, and polarization states,
one arrives at a zero-point energy density for the electromagnetic
field of

$$
W = \int_0^{\omega_c}\rho(\omega)d\omega = \int_0^{\omega_c}
{\hbar\omega^3\over2\pi^2c^3}d\omega,\eqno(1)
$$

\noindent where $\omega_c$ is a postulated cutoff in frequency. In
conventional GRT, this zero-point energy density
must be a source of gravity. This conflicts with astrophysical observations
such as the size, age, and Hubble expansion of the Universe by as much
as a factor of $10^{120}$. Moreover, in addition to the electromagnetic
zero-point energy there is also zero-point energy associated with gluons
and the $W$ and $Z$ vector bosons. From na{\"\i}ve mode counting it would seem
that the gluons should contribute eight times as much zero-point energy as
do the electromagnetic zero-point photons, since there are eight types of
gluons. While this estimate could doubtless be refined with a more
sophisticated examination of the gluon model, it nevertheless seems clear
that the vacuum energy density of gluons must be at least comparable to,
and could quite easily be an order of magnitude or so larger than, 
the vacuum energy density of photons. The massive vector bosons must
likewise provide a contribution of roughly similar scale.
The fields associated with other forces thus exacerbate a
problem that is already difficult when only electromagnetism is considered.

There is no accepted quantum theory of gravity, but ``we might expect on
the basis of studies of weak gravitational waves in general relativity
that the field would also have a ground state energy $\hbar\omega/2$ for
each mode and the two polarization states of the waves."$^{(1)}$ This too
would only compound the problem. 

One possible solution to the dilemma lies in the Dirac 
vacuum. According to theory, the fermion field of virtual quarks,
leptons, and their antiparticles, should have negative energy. If there
were precise pairing of fermions and bosons, as for example results from
supersymmetry, there could be a compensating negative zero-point energy.
Unfortunately, while supersymmetry is often used as a starting point
in modern theoretical investigations, it has neither been proven necessary
nor demonstrated empirically; indeed, the ongoing failure to observe
superpartners for any known particles is a longstanding albeit
minor embarrassment for the theory (see e.g. Ramond 1981$^{(3)}$).

Another approach is more phenomenological in content. It comes from
GRT, though its quantum-field-theoretic 
interpretation is usually connected to the Dirac vacuum approach. 
This technique uses the ``cosmological constant'' of the Einstein
equation to absorb or cancel the effects of an arbitrary energy density.
This will be discussed in more detail in a later section; for now it
is sufficient to note that both of these approaches require cancellation
of opposed densities to an utterly fantastic degree of precision.

One might try taking the position that the zero-point energy must be
merely a mathematical artifact of theory. It is sometimes argued, for
example, that the zero-point energy is merely equivalent to an 
arbitrary additive potential energy constant. Indeed, the potential
energy at the surface of the earth can take on any arbitrary value,
but the falling of an object clearly demonstrates the reality of a
potential energy field, the gradient of which is equal to a force.
No one would argue that there is no such thing as potential energy
simply because it has no well-defined absolute value. Similarly,
gradients of the zero-point energy manifest as measurable Casimir forces,
which indicates the reality of this sea of energy as well. 
Unlike the potential energy, however, the zero-point energy is
not a floating value with no intrinsically defined reference level.
On the contrary, the summation of modes tells us precisely how much
energy each mode must contribute to this field, and that energy density
must be present unless something else in nature conspires to cancel it.

Further arguments for the physical reality of zero-point
fluctuations will also be addressed in later sections. For the
current introductory purposes we may simply observe that
Adler \etal\ $^{(1)}$ summarize the situation thus:

{\narrower\medskip

Quantum field theory predicts without ambiguity that the vacuum has an
energy density many orders of magnitude greater than nuclear density.
Measurement of the Casimir force between conducting plates and
related forces verify that the shift in this energy is real, but 
considerations of gravity in the solar system and in cosmology imply 
stringent upper limits on the magnitude, which are in extreme conflict
with the theoretical estimate, by some hundred orders of magnitude!
Unless one considers an ad hoc constant cancellation term an adequate
explanation then there appears to be a serious conflict between our
concepts of the quantum vacuum and gravity; that is, there is a
vacuum catastrophe.

\medskip}
None of the resolutions to this ``vacuum catastrophe'' suggested above
is entirely satisfactory, but some speculative developments suggest one
more potential alternative. We may consider the possibility that
the electromagnetic and other zero-point fields
really do exist as fundamental theoretical considerations mandate,
but that their zero-point energies do not gravitate because it is the
actions of these fields on matter that generate gravitational forces
(which are mathematically represented by the curving of spacetime). 
The zero-point energies do not gravitate because the zero-point fields
do not, indeed cannot, act upon themselves. The basis of such a zero-point
gravitation theory was conjectured by Sakharov$^{(4)}$ and
Zel'dovich$^{(5)}$ and has undergone a
preliminary development by several authors (see e.g. Adler$^{(6)}$).
More recently, and in consonance with our approach, this
situation appeared in a clearer manner in the attempt of Puthoff.$^{(7)}$

We point to the potential importance and possible direction of a
zero-point gravitation theory, but do not attempt to develop this
ourselves. The principle of equivalence, however, dictates that if
gravitation is an effect traceable to the action of zero-point fields
on matter, then so must the inertia of matter be traceable to zero-point
fields. This approach Woodward and Mahood$^{(8)}$ vehemently find to be
objectionable, treating it as if it were a dangerous new heresy. In their
paper they summarize some connections between gravity and inertia, but
fail to see that this simply establishes relationships that must exist
between the two regardless of whether gravity and inertia are due to
zero-point fields or not. Their arguments about inertia leave the paradox
between quantum theory and gravitation theory as unresolved as ever. 

As alluded to above, the recent work of Haisch,
Rueda, and Puthoff$^{(9)}$, and more recent development by Rueda and
Haisch$^{(10)}$, derives inertial reaction forces from interactions
with the zero-point fluctuations of the quantum vacuum. The contrary
theory of Woodward and Mahood$^{(8)}$ builds on earlier work in gravity
and GRT to suggest that inertia is an extrinsic result of 
interactions with the gravitational field arising from the overall
mass distribution of the cosmos. 

The current analysis consists largely of a
rebuttal to this last reference, and a response to its criticisms. Due
to the frequency of reference, we shall use WM to refer to Woodward and
Mahood$^{(8)}$, HRP to refer to Haisch, Rueda, and Puthoff$^{(9)}$, and RH
to refer to Rueda and Haisch.$^{(10)}$

\bigskip\noindent{\bf 2. CRITIQUE OF GRAVITATIONAL INERTIA}
\bigskip\noindent{2.1 General problems with a gravitational theory of
inertia}
\bigskip\noindent
One of the most striking features of the General Theory of Relativity is
that it essentially banishes the concept of a gravitational force. Gravity,
according to GR, is a distortion of the metric of spacetime. An
object seen by a distant observer to be accelerating in a gravitational
field is, in fact, pursuing a geodesic path appropriate to the spacetime
geometry in its immediate vicinity: no accelerometer mounted on such an
object will detect an acceleration. 

The Principle of Equivalence, adopted by Einstein as a starting point in
the construction of GR, asserts that the state of free-fall one would
encounter in deep space, far from all gravitational sources, is in fact
the same state one encounters while falling freely in a strong gravitational
field.$^{(11)}$
As a corollary of this equivalence, an acceleration relative to the
local free-fall geodesic has the same effects, whatever the local geometry.
Near Earth's surface, for example, geodesic paths accelerate toward Earth's
center. To hold an object at rest relative to Earth's surface, therefore,
requires that it be ``accelerated'' relative to this geodesic by the
application of force; and, by Einstein's original formulation of equivalence,
the effects of this acceleration are indistinguishable from those encountered
in an accelerating reference frame in remote space (see, e.g.
Einstein$^{(12)}$).

In other words, the Principle of Equivalence asserts that gravitational
``forces'' as conventionally measured are inertial reaction forces --
pseudo-forces, as these are sometimes called. We thus see that any attempt
to identify gravity as the source of inertia, within the context of GRT,
suffers from an essential circularity. At the level of ordinary discourse,
this is almost trivially obvious. We consider an extrinsic theory of
inertia which claims that inertial reaction forces are gravitational
forces. But the equivalence principle requires that gravitational forces
are inertial reaction forces, so applying equivalence to the theoretical
claim we see it reduce to the uninformative declaration that inertial
reaction forces are inertial reaction forces. 

To demonstrate that this is not simply linguistic play, let us 
consider the situation with a bit more rigor. The various extrinsic-%
inertia models discussed by WM all have the common feature that they
mandate the appearance of a gravitational field in an accelerated frame
of reference. This is, in fact, quite uncontroversial and in no
way depends on the acceptance of Mach's principle. Traditional, non-Machian
approaches to GRT note that an accelerating reference
frame will see a space-time metric corresponding to a gravitational
field pervading all space. This is quite unsurprising since the 
accelerating observer sees the entire Universe accelerating relative
to itself, and how better to explain this than by a cosmic gravitational
field? The Machian element comes in only when one requires that the
source of this cosmic field should be the overall mass distribution
of the cosmos, rather than an intrinsic property of spacetime. 

Regardless of the source of the cosmic gravitational field, an object
held at rest in it --- that is to say, any massive object sharing the
motion of the accelerating reference frame --- will, of course, exert
weight on whatever agency is holding it at rest. In the reference
frame of the cosmos, on the other hand, the accelerating body is
exerting the expected inertial reaction force on whatever agency is
causing it to accelerate. Have we explained inertia via the cosmic
gravitational field? 

Unfortunately, the standard geometrical approach to GRT
says otherwise. In the presence of a gravitational field, an unconstrained
body must fall freely along a geodesic path. To alter its motion from
this spontaneous condition, one must apply a force to it, creating an
acceleration which will be noted by, for example, any accelerometer 
rigidly mounted on the body. Common experience requires that this will
produce an inertial reaction force as the body's inertia resists this
acceleration. At this point we can identify three alternative explanations
for the inertial reaction:
\item{1.} The inertia is intrinsic to the mass of the body. While this
is consistent with observation it simply postulates inertia without
explaining it. 
\item{2.} The inertia is extrinsic to the mass, being the result of the
interaction of the mass with some non-gravitational field. 
The ZPF-inertia theory of HRP falls into this class.
\item{3.} The inertia is extrinsic to the mass and results from the
interaction of the mass with the apparent gravitational field. This
gravitational explanation of inertia is the one WM are claiming. 

To see how peculiar a theory of the third class above actually is,
let us ask why the inertial reaction force appears at all in this
theory. WM apparently
believe that the presence of a gravitational field in the accelerating
frame is a sufficient explanation: the reaction force is the body's weight
in this field. But why do bodies have weight in a gravitational field?
In the standard formalism of geometrodynamics, gravity is not a force
but a consequence of the local shape of spacetime. ``Weight'' is actually
the inertial reaction force that results from accelerating an object
away from its natural geodesic path. But we are, here, trying to {\it
explain} inertial reaction forces. To say that an inertial reaction force
is the weight resulting from gravity in the accelerated frame explains
nothing in geometrodynamics, because weight is already assumed to be an
inertial reaction force and one is therefore positing inertial reactions
to explain inertial reactions. Therefore, this ``explanation'' of the
origin of inertial reaction forces is circular {\it if} one is operating
in the standard geometrical interpretation of GRT.

It is, of course, possible to abandon this interpretation and presume
that gravity actually does exert forces directly on objects, as in the 
original Newtonian theory. This, unfortunately, introduces a different
circularity. The fact that a gravitational field appears in an accelerating
frame is, as noted above, true in any formulation of GRT,
Machian or not, and remains true whether inertia is intrinsic or extrinsic.
The gravitational-inertia theory wishes to assert that this gravitational
field is the cause of the inertial reaction force. But this is the same
as the assumption that gravitational fields exert forces; we cannot claim
to have explained inertia in this formalism when we incorporate our
desired conclusion into the initial assumptions. 

This would appear to be a very general problem with efforts to find a
gravitational origin for inertia in the standard, geometrodynamic 
interpretation of GRT. There are, of course, ways around
this. An argument by Sciama$^{(13)}$, for example, finds a reaction
force arising from a ``gravito-magnetic'' reaction with a presumed
gravitational vector potential. It is, however, well worth noting that
Sciama's argument is based on analogizing gravitation to electromagnetism,
in the weak-field limit of GR. In this weak-field limit one typically
does not work explicitly with the geometrical consequences of metric
distortion, but rather represents interactions in terms of potentials
and forces. The circularity noted above disappears, but with it the
conceptual parsimony of GR.
Indeed, as WM themselves assert (their section 3.2),
Sciama's argument was originally conceived as a refutation of GRT. 

{\it General relativity, in reducing gravity to a consequence of geometry,
offers a very hostile background to a gravitational theory of extrinsic
inertia. GR shows how mass distorts spacetime, and allows one to calculate
the trajectories unconstrained bodies will follow in the resulting
distorted spacetime. It does not explain why a body, constrained by
non-gravitational forces to travel on some trajectory that is not a
geodesic, exerts an inertial reaction force proportional to its mass.}

This is, of course, a trivial non-mystery if one na{\"\i}vely 
presumes inertia to be
intrinsic to mass. The attempt, however, to construct a gravitational
theory of extrinsic inertia within geometrodynamics seems doomed to
circularity. 

\bigskip\noindent
{2.2 Specific problems with WM argument}
\bigskip\noindent
In fairness to WM they do seem aware, to a certain extent, of the
circularity problem. At the end of their section 3.4 they devote
a paragraph to an attempt to address it. Unfortunately, they dilute
and weaken
their argument by attempting to portray the circularity argument as
a defense of ZPF-inertia theory, which it is not. Indeed, it would
seem that the WM response to the the circularity argument consists
mainly of the complaint that ZPF theories do not successfully explain
inertia either, which even if it were the case 
is irrelevant to the failure of
gravitationally based theories to do so. One should bear in mind
that the default explanation of inertia, currently highly favored
by Ockham's Razor as the least hypothesis, is that inertia is intrinsic
to mass. Various important elements of physical theory, such as the
conservation of momentum, which flow quite naturally from a theory
of intrinsic inertia, require complicated supporting arguments or may
even be violated in a theory of extrinsic inertia. (It is worth
noting that one of the authors of WM has in fact published articles --- and
obtained a U. S. Patent$^{(14)}$ --- demonstrating ways in which a theory of
extrinsic gravitational inertia allows local violations of momentum
conservation.$^{(15)}$ While one might hope, and indeed the same
papers claim, that momentum is still conserved globally, this is actually a
meaningless assertion in the Machian perspective of this theory.)

In their section 3.2 WM make the peculiar claim that ``GRT dictates that 
inertia is gravitationally induced irrespective of whether cosmic
matter density is critical or not." This claim is odd, because it
seems to be supported only by the assertion that in Robertson-Walker
cosmologies the local metric is determined solely by the distribution
of material sources within the current horizon. While this claim
is true, it does not address the relationship between critical
density and gravitational inertia. All of the arguments employed by
WM require a specific value for the total gravitational potential
$\phi$ in order for inertial reaction forces to behave properly.
This depends on the cosmic mass density $\rho$ in a Robertson-Walker
cosmology. While WM's demonstration that sources outside the horizon
may safely be ignored is valid and useful, it falls badly short of
explaining why the actual density of sources {\it inside} the horizon
can also be ignored in declaring that physics is Machian and inertia
results from gravity.

In section 3.3 WM provide a general discussion of the relation between
Mach's principle and GRT. In the current context this is notable mostly
for its complete omission of results suggesting that GRT is not only not a
Machian theory, but in fact incompatible with Mach's principle. For
example, the Lense-Thirring precession is often touted as an example
of the ``Machian" dragging of inertial frames by a rotating mass, but
recent work by Rindler$^{(16)}$ demonstrates that the equatorial 
Lense-Thirring effect is inconsistent with a Machian formulation.
Granted, the Lense-Thirring rotation is such a minute effect that
it has not been empirically tested, but it is an unambiguous prediction
of GRT: to have an anti-Machian effect emerge from GRT
impedes the joint claim of WM that GRT is the correct theory of gravity
and that the Universe is Machian.

WM go on in section
3.4 to discuss an argument by Nordtvedt$^{(17)}$ concerning frame dragging
in translational acceleration. They present as their eq. 3.7 the
relation:

$$\delta{\bf a} = (4\phi/c^2){\bf a},\eqno(2)$$

\noindent which relates the induced (frame-dragging) acceleration 
$\delta{\bf a}$ to the acceleration ${\bf a}$ of the accelerated mass
and the gravitational potential $\phi$ induced by that same mass. 
They point out that if $4\phi = c^2$, then $\delta{\bf a} = {\bf a}$
and all inertial frames are dragged rigidly along with the inducing
body. If one regards the universe at large as Nordtvedt's inducing
body, and presumes that it has the appropriate value of $\phi$
throughout, then any hypothetical acceleration of the universe
would necessarily drag along all inertial frames; an alternative
way of expressing this is to say that the bulk mass distribution of
the cosmos defines which frames are inertial. So far this would
appear to be an excellent demonstration of Mach's principle.

As a possible quibble we note that for $\phi > c^2/4$ the ``frame
dragging" acceleration is {\it greater} than the acceleration
of the inducing body, a bizarre result that seems very difficult
to attribute to frame dragging. In fact, as WM acknowledge, 
Nordtvedt's derivation is of linear order in the mass, and is 
therefore of questionable validity for the large values of $\phi$ 
they wish to apply. But this ranks only as a quibble, because the
problem of inertia has not been addressed at all. Even if one,
implausibly, stipulates the validity of eq.~2 over all $\phi$,
one has merely identified which states of motion are inertial
reference frames: no explanation has been offered for the appearance
of inertial reaction forces in non-inertial frames. We are once
again facing the circularity problem of the previous section,
with no progress toward an explanation. As noted above, WM have
not successfully addressed this problem anywhere in their discussion
of gravitational inertia.

The next difficulty in WM is perhaps best introduced by quoting
their own argument, noting that $\phi$ is their symbol for total
gravitational potential as in eq.~2 above.

{\narrower\medskip

Since the locally measured value of $\phi$ must be an
invariant to preserve the principle of relativity, one might think
that the gradient of the gravitational potential must vanish 
everywhere. Accordingly, it would seem that no local gravitational
fields should exist. But the gradient of a locally measured invariant
need not vanish if it is not a {\it global} invariant. The total 
gravitational potential is not a global invariant. As a result, the 
``coordinate" value of the gravitational potential in some frame of
reference may vary from point to point, notwithstanding that the 
numerical value measured at each point is the same everywhere. And 
the gradient of the potential in these coordinates may be non-vanishing.
As a familiar example of this sort of behavior we point to the vacuum
speed of light --- a locally measured invariant --- in the presence of
a gravitational field. As is well known, the speed of light in intense
gravitational fields measured by {\it non-local} observers (that is, the
``coordinate" speed of light) is often markedly different from the
locally measured value. And for these non-local observers, the speed
of light in general will have a non-vanishing gradient in their
coordinates. (WM, section 4.2, excerpt from final paragraph.)

\medskip}
Clever as this argument and analogy may seem, it introduces a new
paradox worse than the one they seek to evade. The speed of light in vacuum
is deeply embedded in relativistic kinematics. If a given coordinate
system measures an altered value of $c$ in some remote regions, it will
also note distortions in lengths and time intervals in those regions
such that it will expect an observer in that region to find the standard
local value for $c$. The potential $\phi$, on the other hand, is a dynamic
variable, not a kinematic one. Where $c$ appears in such fundamental and
inescapable relations as the velocity-addition rule, $\phi$ is merely
a potential; its value dictates how specific objects will move, not the
nature of motion itself. 

Let us posit the WM scheme of a locally
invariant $\phi$ that is nevertheless observed to vary and have a gradient
in certain reference frames. The quantity $\phi$ is, by definition, a
gravitational potential: $m_g\phi$ is the gravitational potential energy
of an object with gravitational mass $m_g$. The value of $\phi$ used in
computing this quantity is, of course, the local value at the current
position of the object. If $\phi$ is a local invariant, no object can
change its gravitational potential energy by moving from one location
to another. A distant observer, seeing an object move from a region with
potential $\phi_0$ to a region at a different $\phi_1$, would expect to
see its kinetic energy change by the quantity $m_g(\phi_0 - \phi_1)$.
A comoving observer, in contrast, observing that the gravitational 
potential energy is $m_g\phi$ at both locations, does not expect any
change in the relative velocity of the object with 
respect to the rest of the cosmos.
These conflicting expectations cannot be reconciled.

As if the above problems were not enough, this new perspective on 
$\phi$ shows that the Nordtvedt frame-dragging effect of eq.~2
above is, rather than a support of the WM inertia
theory, absolutely fatal to it. If $\phi$ is a locally measured
invariant due to the action of the entire cosmos, no local concentration 
of matter can affect $\phi$, which leads to the startling conclusion
that {\it no body smaller than the Universe as a whole can produce
any frame dragging effects whatsoever!} WM require this locally
invariant character for $\phi$ in order to avoid having inertia behave
unacceptably (that is, in a manner contrary to long-established observation)
in the vicinity of gravitating masses. Yet the price of this local
invariance is the disappearance of all local frame-dragging effects.
And, again as WM themselves point out, Nordtvedt's frame-dragging 
effect is necessary for such quotidian phenomena as planetary
orbits to display the proper invariance under arbitrary choices
of coordinates. 

In their section 4.3 WM refer to a ``stronger version'' of Mach's
Principle, in which ``...{\it mass itself} arises from the gravitational
action of the distant matter in the universe on local objects ---
mass is just the total gravitational potential energy a body possesses."
Unfortunately this does not work, at least not in the all-encompassing
sense that WM seem to have in mind. In order to establish the gravitational
potential energy of a body, one must have at least one kind of mass,
the gravitational mass $m_g$, as a preexisting quantity, so that $m_g\phi$
gives the total gravitational potential energy. This version of Mach's
principle would allow one to derive the energetic content of mass 
and explain why $E/c^2\equiv m_g$, but does not quite explain mass
itself {\it ex nihilo} as WM appear to be claiming. 

While certain other parts of WM's explication of gravitational inertia
are flawed, these closely involve their criticisms of ZPF theories, and
so discussion of them is better deferred to the next section.

\bigskip\noindent{\bf 3. CRITICISMS OF ZPF: ERRORS AND CORRECTIONS}
\bigskip\noindent
WM raise numerous criticisms, both of the notion of quantum zero-point
fluctuations and of the specific HRP theory of extrinsic inertia based
on interactions with ZPF. Most of these are severely
flawed. Before dealing with the WM criticisms in detail, it is worth
noting that the strongest criticism is not one that they raise
explicitly, though it is implied by certain of their other arguments.
The exact identity between the inertial mass which resists accelerations,
the gravitational mass which acts as a source term in the Einstein field
equation, and the energetic-content mass $E/c^2$ follows quite naturally
in simplistic intrinsic-inertia theories. 
It needs careful attention, though, in any theory
of extrinsic inertia, and the ZPF-inertia theory put forward in HRP 
is not yet able to account for this identity. Since the ZPF-inertia 
theory is still in its early stages of development, this should not
be considered either surprising, or a refutation of the theory. 

The various points raised in WM actually address two distinct issues,
the physical reality of ZPF and the theory that ZPF interactions are
the cause of inertial reaction forces. Obviously the former issue is
logically prior to the latter; it is also empirically of greater
consequence, since the existence of ZPF-driven effects such as the Casimir
force and the Lamb shift have been confirmed experimentally. Some
alternative explanation for them must be found if we wish to keep our
theories in consonance with reality. We will therefore address the
existence of the ZPF first.

\bigskip\noindent
{3.1 Elementary theoretical justification}
\bigskip\noindent
The Introduction above, in explaining the $\approx120$ order-of-magnitude
discrepancy that motivates the search for a ZPF-inertia theory,
already provided several strong arguments for considering the ZPF
physically real. One further argument worthy of consideration,
however, emerges
from experiments in cavity quantum electrodynamics
involving suppression of spontaneous emission. As Haroche and Ramond
explain$^{(18)}$:

{\narrower\medskip

These experiments indicate a counterintuitive phenomenon that might
be called ``no-photon interference." In short, the cavity prevents an atom
from emitting a photon because that photon would have interfered 
destructively with itself had it ever existed. But this begs a 
philosophical question: How can the photon ``know," even before being
emitted, whether the cavity is the right or wrong size?

\medskip}
 
\noindent The answer is that spontaneous emission can be interpreted as 
stimulated emission by the ZPF, and that, as in the Casimir force
experiments, ZPF modes can be suppressed, resulting in no vacuum-%
stimulated emission, and hence no ``spontaneous" emission.$^{(19)}$

\bigskip\noindent
3.2 The cosmological constant problem
\bigskip\noindent
WM object that ``$\ldots$if the ZPF really did exist, the gravitational
effect of the energy resident in it would curl up the universe into a
minute ball" (section 2.2, WM). This, of course, is precisely the
vacuum catastrophe problem discussed in detail in the Introduction. When various
solutions to that quandary were being discussed, it was pointed out that
several of them require an implausibly precise cancellation between the
ZPF energy density and other physical factors. However, one of those
theoretical devices --- the cosmological constant 
--- suffers a fine-tuning problem, whether or not it is invoked to avoid
the vacuum catastrophe. The general form of the Einstein field
equation,

$$R_{\mu\nu} - {1\over2}g_{\mu\nu}R +\Lambda g_{\mu\nu}
= -{8\pi G\over c^4} T_{\mu\nu},\eqno(3)$$

\noindent
includes an arbitrary ``cosmological" constant $\Lambda$. 
This term can absorb any contribution from a uniform density such as
the vacuum energy. As noted in the Introduction, actually matching
the ZPF energy density would be a feat of remarkable precision. The
fine-tuning problem persists even if one assumes that something else
averts the vacuum catastrophe, because observational astronomy 
increasingly favors a cosmology with a small
nonzero value of $\Lambda$. Unfortunately, field-theoretic 
considerations suggest that ``natural'' values of $\Lambda$ should
be either exactly zero, or else correspond to an energy density 
(positive or negative) on the rough order of one Planck mass per Planck
volume. We are thus confronted with a fine-tuning problem for $\Lambda$
whether or not we wish to use it to resolve the ZPF energy density
problem. 

\bigskip\noindent
3.3 Local fluctuations versus nonlocal interactions
\bigskip\noindent
WM point out that ``$\ldots$ {\it any local fluctuational explanation
can be reinterpreted as a non-local, retarded/ advanced interaction
with distant matter.}'' (Section 4.4, emphasis in the original.) This
may very well be true, but it can scarcely be taken as support for
their thesis. Insofar as there is a consensus in the physics community
on the issue of nonlocality, it would seem to be that nonlocality is
to be avoided at almost any cost. WM refer to the well-established
``nonlocal" interactions of quantum mechanics (earlier in their section
4.4 than the above quote) in an attempt to justify their preference
for a nonlocal explanation of ZPF-driven effects. Unfortunately, what
quantum mechanics refutes is not locality but the conjunction of
locality with some aspects of objective realism. (The minimal part
of realism that must be rejected has been labeled ``contrafactual
definiteness," the notion that it is meaningful to discuss the
potential outcomes of experiments that might have been performed
but in fact were not.) By observation, most physicists confronted
with the failure of local realism prefer to abandon some aspect of
realism rather than some part of locality.$^{(20)}$

Other justifications WM present for preferring a theory that mixes
retarded and advanced waves are the utility of Feynman-Wheeler
absorber theory and the recent proposal of Cramer's ``transactional
interpretation" of quantum mechanics. Remarkable though the Feynman-%
Wheeler theory is, we should not lose sight of the fact that it is
one of several formalisms that all account successfully for the
non-observation of advanced waves.  The ``transactional
interpretation," on the other hand, is {\it by construction} devoid
of empirical content: all philosophical interpretations of quantum
mechanics of necessity agree with all empirical predictions of QM
and therefore permit no empirical preference for one over another.
One's choice of QM interpretation is therefore a matter for philosophical
aesthetics rather than scientific judgement. 

Contrary to the claims of WM, standard relativity theory in no way
demands the ``radical timelessness" they advocate. At least, it does
not do so as long as nonlocal interactions are kept from contaminating
the theory. In a conventional relativistic world without nonlocality,
time proceeds in a well-ordered fashion along every timelike worldline.
The inability of observers in different states of motion to agree on
the relative ordering of remote, spacelike-separated events is 
irrelevant; this ambiguity can never lead to causal confusion or lead
to ``future" events affecting the ``past." Essentially, this is because
the conventional interpretation of relativity replaces the traditional
view of past, present and future with a four-part division of reality.
From any given event, the ``future" encompasses everything in the
future light cone, the ``past'' the entire contents of the past light
cone. ``Now," which a Newtonian physicist could conceptualize as a
shared instant of simultaneity encompassing all space, has shrunk to
the single space-time point of the event under consideration. And
the rest of the universe is in a region commonly dubbed ``elsewhere,"
a constellation of space-time events that can neither affect nor be
affected by the event under consideration in any way. So long as all
interactions are local, the potentially inconsistent time-ordering
of events ``elsewhere" can never lead to the slightest confusion
between events in the past and events in the future, nor allow the
latter to affect the former. 

This of course breaks down if one admits of nonlocal interactions.
By means of a nonlocal connection an event in the future light-cone
can send a signal to an event ``elsewhere," and cause a returning
nonlocal signal to arrive at an event in the past. This should make
it clear that it is not relativity, but relativity plus nonlocality,
which demands the radical timelessness and its ``very strange
consequences" advocated by WM.

Having addressed WM's primary arguments against the physical reality
of ZPF in general, we now turn to their arguments against the HRP
theory of ZPF as the origin of inertia.

\bigskip\noindent
3.4 A Sketch of HRP's and RH's Claims
\bigskip\noindent
In the discussion by this name in their section 2.1, WM, in order to
criticize the arguments of HRP and RH, present a simplified argument that
in their terminology is intended to uncover ``the crux of the whole
business." A simplified argument which still contained the essential
physical ingredients of the calculation would be a useful pedagogical
as well as conceptual excercise. It must, however, remain physically
accurate. Unfortunately this is not the case with the presentation of WM,
which, despite their claim of ``accurate formalism'', is both misleading
and erroneous.

Before discussing this presentation in detail, however, it seems desirable
to clarify the motivations two of the current authors (AR and BH) had for
producing the HRP and RH papers. The HRP paper involved a detailed 
calculation of the behavior of a Planck oscillator pushed by an external
agent to move under uniform proper acceleration (so-called hyperbolic
motion). In spite of some simplifying assumptions and a few fairly
reasonable approximations, the mathematical development of the HRP article
came out to be quite complex. The inertia effect was clearly obtained but
assessment of the calculations and of the argument was challenging. 
It was not clear whether there was something in the vacuum, as viewed
from an observer comoving with an accelerated frame, that could produce
the effect predicted in HRP. Calculations in QED and QFT for a detector
accelerated in a {\it scalar} vacuum field did not seem to find any
anisotropy in the scalar field even though the well-known
Unruh-Davies thermal background was predicted to occur.$^{(21)}$  It was
necessary to check if the {\it vector} nature of the electromagnetic ZPF
(as opposed to a scalar field) would produce the expected anisotropy in the
vacuum background from the viewpoint of such a uniformly accelerated
observer.

This problem was attacked and a confirmatory result emerged from the 
calculations. After approaching the problem in four different ways,
as detailed in RH, it was in 
all four ways clearly found that an anisotropy appeared in the ZPF 
Poynting vector and hence that an anisotropy appeared in the flux of
momentum density. More than that, the anisotropy in the Poynting vector
was of the precise form to produce a radiation pressure opposite to the
acceleration and proportional to it in the subrelativistic case, and
also extended properly to the standard relativistic form of the inertial
reaction 4-force at large speeds. 

In their section 2.1 WM attempted to do two things, both of which were 
commendable in principle. First, they tried to present a simplified
pedagogical view that would clearly illustrate the physics of the
situation analyzed in the calculations presented in HRP and RH. Second,
they attempted to relate the analysis of RH to that of HRP so that
the physics of the inherent connection could easily be seen. We must
report, however, that they were unfortunately unsuccessful in both of
these endeavors. The main point of this part of their presentation
in this respect was to replace eqs. (26) to (28) of HRP by the
very simple proportionality relationship between the electric field
${\bf E}_{zp}$ and the velocity ${\bf v}$ of vibration of the subparticle
component in the instantaneous inertial frame of reference at particle
proper time $\tau$, in the form of WM eq. 2.1:

$$e{\bf E}_{zp} = k {\bf v}.\eqno(3)$$

\noindent This enormous simplification had the following consequences:
\item{(i)} All ${\bf E}$-field frequency components and all components in
all directions seemed to contribute with the same weight to the instantaneous
velocity of the subparticle, contrary to the facts. 
\item{(ii)} All those contributions appeared to come exactly in phase,
contrary to the facts.
\item{(iii)} As a consequence of (i) and (ii) we get the physically very
surprising feature that the electric field force was proportional to
the velocity. (This might be called Aristotelian physics.) But we know this
cannot happen unless energy is not conserved, or more precisely, unless
energy goes to degrees of freedom that have not been accounted for in
detail, as happens with a thermal reservoir. In reality the Planck
oscillators interact with the ZPF in a dissipationless manner,
so the dissipative force in the WM analysis is both inaccurate and
misleading.

After such a disastrous start in the first equation, it is tempting
to simply discard the entirety of WM's subsequent argument. In
particular, since WM eq. 2.3 depends on the inaccurate 2.1, it is
itself invalid, and all conclusions drawn from it are suspect. However,
there are additional and independent errors in the WM analysis which
merit separate comment. 

To reprise briefly the development of the HRP/RH argument given above: 
The inertialike
reaction force appearing at the end of the HRP derivation implies the
necessary existence of an anisotropy in the accelerated ZPF. However,
earlier work in vacuum scalar fields found no such anisotropy. 
RH therefore investigated the existence of such anisotropy
in vector fields, and found a net Poynting vector in accelerated 
vector ZPF by four separate lines of argument.

However, in RH no details on the particle were used since the analysis
concentraed on the fields. The Poynting vector appears in the accelerated
ZPF regardless of any entity that may interact with it. That interaction
was introduced only at the end, in the form of a normalizing function
$\eta(\omega)$ that quantified the momentum density passed to the
accelerated object at every frequency. In contrast, the original HRP
analysis modeled this interaction in great detail. In this case the
Einstein-Hopf model was used, which implied only a first-order iterative
solution and hence some degree of approximation. The considerable
difference in methods between RH and HRP is the reason for the difference
in appearance of the inertial mass expressions in RH and HRP. It seems
likely that to derive the RH form from the expressions of HRP one would
have had to pursue an iterative solution to many orders, going far beyond
the Einstein-Hopf approximation. 

The discussion presented by WM contrasts with the detailed analysis
done in RH and HRP. For a serious discussion of the technical aspects
of HRP (and to a lesser extent RH) we prospectively refer the interested 
reader to works presently 
in progress by Cole and Rueda, and by Cole.$^{(22)}$

\bigskip\noindent
3.5 The problem of representing the accelerating body
\bigskip\noindent
Aside from the general flaws of WM section 2.1 noted above, we note
that their simplified model includes the assumption that the ``oscillator''
interacting with the ZPF is in fact an elementary point charge.
This is problematic. A point charge in classical theory has infinite
self-energy, leading to some question of whether it is legitimate to
deal with such objects except as an approximation good for long wavelengths
and modest accelerations. This, unfortunately, is the exact opposite of the
regime crucial to the ZPF-inertia theory. The empirical verification of
quarks (or leptons) as pointlike extends only to length scales orders
of magnitude longer than the wavelengths important to either the 
HRP or RH derivations. The representation of the particle/radiation
interaction, in the one case by a generalized damping coefficent 
$\Gamma$, in the other by an unspecified interaction function $\eta(
\omega)$, seems appropriately cautious at our current level of ignorance.

\bigskip\noindent
3.6 The bare mass problem
\bigskip\noindent
In the discussion subsequent to their eq. 2.8 WM discuss the
apparent circularity of using $\Gamma=2e^2/3m_0c^3$, with a
contribution from a ``bare'' mass $m_0$ with presumed inertial 
effects, in the HRP derivation that purports to identify the source of
inertial mass. This is a valid criticism, which suggests that a
reworking of the formalism is desirable. In fact the later work of
RH presents such a reworking, with no reference to unobservable 
``bare'' masses. 

\bigskip\noindent
3.7 Quark and hadron masses
\bigskip\noindent
The extended discussion WM conduct in their section 2.2 on this
issue implies the general mass-equivalence problem which, as noted 
above, is a valid concern and an unmet challenge for the ZPF-inertia
theory. However, the specific points made by WM are, as they 
themselves point out, largely answered by HRP; and their rebuttal
of this answer appears to misunderstand it. As is clearly indicated
in the text WM choose to quote, the authors explicitly propose
a revised formalism in which the interaction is assumed to be
dominated by a resonance frequency $\omega_0$, determined by
the particle dynamics, rather than the ZPF cutoff frequency $\omega_c$.
WM respond to this proposed model by asserting:

{\narrower\medskip

Well, $\omega_c$ isn't a ``resonance" frequency. It is the upper 
limit in the integration over the frequency spectrum of the ZPF, and if
that limit is not imposed, the result of that integration, and the 
inertial mass of the particle, is infinite irrespective of any resonances
that may be present at finite frequencies.
Remember, the spectral energy density of the ZPF
goes as $\omega^3$, so invoking a ``low" frequency resonance will
not suppress the cutoff unless the cutoff is assumed to lie quite
close to the resonance frequency.

\medskip}

But this counterargument is clearly without merit. Any resonant
phenomenon with a frequency response that falls off sharply enough
for $\omega >\omega_0$ will have a converging and therefore finite
integral in the reaction-force calculation. And the criterion for
``sharply enough" is much less stringent than WM seem to imagine.

HRP present, in their eq. (3), the
spectral energy density of the ZPF in an accelerated frame.
We reproduce this equation (aside from a common factor $d\omega$
on both sides) here: 

$$\rho(\omega) = \left[{\omega^2\over\pi^2c^3}\right]
\left[1+\left({a\over\omega c}\right)^2\right]
\left({\hbar\omega\over2}+{\hbar\omega\over e^{2\pi c\omega/a}-1}\right).
\eqno(4)$$

We can see that there are four terms when this expression is multiplied
out. One has $\omega^3$ spectral dependence and is in fact the
unaltered $\hbar\omega^3/2\pi^2c^3$ ZPF spectrum itself. This means
that an accelerated reference frame contains the same ZPF as in an inertial
frame, plus three new components. Of these three, one is the
thermal bath identified with the Davies-Unruh effect, one is not
thermal but is, like thermal radiation, suppressed as $e^{-\omega}$
for large $\omega$, and the third and last has a spectral dependence
of $\omega$. It is this last term, varying as $\omega$, not $\omega^3$,
which HRP propose as the source of the reaction force in
their discussion consequent to this formula. 

If we assume then that the radiation term responsible for the reaction force 
has a frequency dependence of
$\omega$, it follows naturally that any resonance centered on a
frequency $\omega_0$ will have a finite total reaction force
integral, even in the limit $\omega_c \rightarrow\infty$, so 
long as its frequency response falls off faster than $\omega^{-2}$
for $\omega\gg\omega_0$. Even if we retain the assumption that the
inertial reaction force derives from the full ZPF spectrum with
its $\omega^3$ energy density, a resonance falling off faster than
$\omega^{-4}$ will remain finite regardless of cutoff.

This point incidentally answers the objection WM raise to the notion
of changes in resonance being responsible for the inertial mass of
a proton. They object that, since the scale of a proton is 20 orders
of magnitude larger than the Planck length, resonances due to the 
proton's structure are 20 orders of magnitude lower in frequency than
the cutoff $\omega_c$. But we have just seen that the cutoff frequncy
is irrelevant. The difference between the electron mass of .511 MeV,
the quark mass of $\approx$10 MeV, and the hadron mass of $\approx$940
MEV can, at least in principle be accomodated by particle-specific 
resonances. These would almost certainly be different for a bound
triplet of particles than some linear summation of individual
resonances for three unbound particles. 

If the electron has a resonant frequency $\omega_e$, we must presume
that a ``free" quark has a resonant frequency $\omega_q\approx20\omega_e$
to account for their mass difference. The term ``free" is used loosely,
since of course color confinement demands that there really is no such
thing as a free quark. What is commonly reported as quark mass is inferred 
from high-energy collisions between various sorts of projectiles and
components within hadrons; the phenomenon of ``asymptotic freedom"
in quantum chromodynamics means that in such high-energy interactions
the quark is little constrained by the color force and behaves almost
as a free particle. On the other hand, in the low-energy state of an
unexcited proton or neutron, the quarks are presumably distributed as
widely as is consistent with color confinement --- if they were more
closely clustered than necessary, the resulting momentum uncertainty
would equate to excess internal energy which would swiftly be emitted
as gamma rays or possibly other particles. In the normal conditions
within a proton or neutron, then, we would expect quarks to be
strongly bound by the color force; and thus, there is plausible
justification in principle for
their resonance at a frequency $\omega_p\approx30\omega_q$.

Moreover, a less strained justification is available. The HRP derivation
deals only with EM vacuum fluctuations, as does the RH analysis. WM,
in castigating an implied model of gluons as vast clouds of charged dust
(to produce EM-ZPF reaction effects), overlook the fact that gluons,
too, have a vacuum fluctuation spectrum. This fact was pointed out in
the introductory discussion of the vacuum catastrophe problem; it does
not disappear merely because we are examining a different consequence
of ZPF effects. Electrons, being colorless, do not interact at all with
gluon fluctuations. We must expect, however, that colored quarks do so
quite strongly. If the ZPF-inertia theory gives the correct explanation
of inertial reactions, therefore, all color-bearing particles must
experience intense inertial reaction effects from a field orders of
magnitude stronger than electromagnetism. 

We may note in passing that this disposes of another WM criticism,
that elementary particles do not show inertial masses proportional
to the squared particle charge $e^2$. Since both $e^2$ and $\omega_0$ 
are factors in the inertial
mass, and a general theory for $\omega_0$ values is not yet available,
we cannot expect $m_i\propto e^2$ to hold between different particles
at even a heuristic level. Nor does the $e^2$ argument pay the slightest
attention to the interaction of particles with fields other than the
electromagnetic.

\bigskip\noindent
{\bf 4. DISCUSSION AND CONCLUSIONS}
\bigskip\noindent
In reviewing the arguments of Woodward and Mahood (1999), the following
conclusions can clearly be seen:

\item{1.} Within the standard geometrical interpretation of general
relativity, any attempt to identify gravity as the source of inertial
reaction forces can succeed only by postulating the thesis it purports
to prove. Such arguments can therefore be dismissed as circular.

\item{2.} While one can construct a gravitational theory for inertial
reaction forces, as in the case of Sciama's 1953 theory, such theories
are necessarily theories of explicit forces coupled to a source $m_g$,
and therefore are quite distinct from the geometrical theory we know
as general relativity.

\item{3.} The particular gravitational-inertia theory propounded by WM
suffers a consistency problem in the handling of $\phi$ as a quantity
that (a) acts as a potential, (b) has a gradient, and (c) is a locally
measured invariant. These three properties prove to be mutually 
incompatible. 

\item{4.} The advocacy of WM for the philosophy of ``radical timelessness"
is, contrary to their own assertion, not a consequence of relativity
but a consequence of their acceptance of nonlocal interactions in a
relativistic framework. 

\item{5.} The arguments of WM against the existence of quantum zero-point
fluctuations are deeply flawed, being based in one case on a misunderstanding
of the cosmological constant problem and in the second case on a willingness
to adopt nonlocal interactions in a way which most working physicists would
find unacceptable. 

\item{6.} The arguments of WM against the HRP theory of extrinsic inertia
arising from interactions with the ZPF make it clear that WM have 
misunderstood almost every important point of the argument. Their
arguments are in most cases invalid, in
some cases useful criticisms pointing to ways in which the theory needs
to be strengthened and improved. In no case whatever 
do they constitute actual refutations.

\medskip Finally, we should note that among the possible theories 
of inertia the most plausible current contender, albeit also the
least informative, remains the simplest:  That inertia
is inherent in mass. No theory of extrinsic inertia yet proposed has
been able successfully to reproduce all of the observed phenomena which are
trivial consequences of this simple premise. The alternative theories of
extrinsic inertia require considerable further development before they
can practically replace the standard interpretation of inertial reaction 
forces which has been thoroughly successful since the days of Newton.

\bigskip\noindent
{\bf ACKNOWLEDGEMENTS}
\bigskip\noindent
B.H. and A.R. acknowledge support of this work by NASA contract NASW-5050.

\bigskip\noindent{\bf REFERENCES}
\bigskip

\item{(1)} R.~J.~Adler, B.~Casey and O.~C.~Jacob, {\it Am. J. Phys.} {\bf 63}, 720 (1995).

\item{(2)} R. Loudon, {\it The Quantum Theory of Light,}
Clarendon Press, Oxford, (1982).

\item{(3)} P. Ramond, {\it Field Theory --- A Modern Primer.}
Beryaunka, Menlo Park CA USA, pp. 55 ff. (1981).

\item{(4)} A.~D.~Sakharov, {\it Dokl. Acad. Nauk SSSR} {\bf 177}, 70 (1968);
translated in Sov. Phys. Dokl {\bf 12}, 1040 (1968).

\item{(5)} Ya.~B.~Zel'dovich, {\it Usp. Fiz. Nauk.} {\bf 95}, 209 (1968);
translated in Sov. Phys. Usp. {\bf 11} (3), 381 (1968).

\item{(6)} S.~L.~Adler,  {\it Rev. Mod. Phys.} {\bf 54}, 729 (1982).

\item{(7)} H. E. Puthoff, {\it Phys. Rev. A} {\bf 39}, 2333 (1989).

\item{(8)} J. F. Woodward and T. Mahood, {\it Found. Physics}, in press.
(1999).

\item{(9)} B. Haisch, A. Rueda and H.E. Puthoff, {\it Phys. Rev. A} {\bf
49}, 678 (1994).

\item{(10)} A. Rueda and B. Haisch, {\it Physics Letters A} {\bf 240}, 115 (1998); A. Rueda and
B. Haisch, {\it Found. Physics} {\bf 28}, 1057 (1998).

\item{(11)} Strictly speaking this is true only for translational motions
of spinless point particles. 
Since nothing in this discussion of inertial reactions
depends either on physically extended bodies or on the presence or absence of
tidal forces, this caveat is irrelevant to the remainder of the analysis.
We likewise ignore the general relativistic spin-orbit interaction, 
since in any gravitational field there exist trajectories for which it 
vanishes.

\item{(12)} A. Einstein, {\it Ann. der Phys.} {\bf 49}, p. 769 (1916) as
quoted in H. C. Ohanian and R. Ruffini {\it Gravitation and Spacetime.}
Second Edition. W.~W.~Norton \& Company, New York, London, p. 53 (1994).

\item{(13)} D.~W.~Sciama, {\it Mon. Not. Roy. Astron. Soc.} {\bf 113}, 34 (1953).

\item{(14)} U. S. Patent No. 5,280,864, ``Method for Transiently Altering the Mass of Objects
to Facilitate Their Transport or Chage Their Stationary Apparent Weights,'' Inventor: James F.
Woodward.

\item{(15)} James F. Woodward, 1997. ``Mach's Principle and Impulse Engines:
Toward a Viable Physics of Star Trek?" Presentation to NASA Breakthrough
Propulsion Physics Workshop, Cleveland, Ohio, August 12-14, 1997, Proc. NASA Breakthrough
Propulsion Physics Workshop, NASA/CP-1999-208694, p. xx (1999).

\item{(16)} W. Rindler, {\it Physics Letters A} {\bf 187}, 236 (1994).

\item{(17)} K. Nordtvedt, {\it Int. J. Theor. Phys.} {\bf 27}, 1395 (1988).

\item{(18)} S. Haroche and J. M. Raimond, {\it Scientific American} {\bf 268},
No. 4, 54 (1993). 

\item{(19)} W. McCrea, {\it Q. J. Royal Astro. Soc.}, {\bf 27}, 137 (1986).

\item{(20)} J. T.~Cushing and E. McMullin (eds.), {\it Philosophical
Consequences of Quantum Theory.} University of Notre Dame Press, 
Notre Dame, Indiana (1989).

\item{(21)} J.R. Letaw, {\it Phys. Rev. D} {\bf 23}, 1709 (1981);
P.G. Grove and A.C. Ottewill,  {\it Class. Quantum Grav.} {\bf 2}, 373 (1985).

\item{(22)} D.C. Cole and A. Rueda, 1999 (in preparation) and D.C. Cole, 1999 (in preparation).

\vfill\eject\end